\documentstyle[preprint,aps]{revtex}

\newcommand{\mathR}{{\rm I\! R}}                

\begin{document}
\draft
\title{\hfill {\rm IMPERIAL/TP/95--96/2}\\[0.8cm]
	Perennials and the Group-Theoretical Quantization \\
	of a Parametrized Scalar Field on a Curved Background}
\author{P.~H\'{a}j\'{\i}\v{c}ek\thanks{email: hajicek@butp.unibe.ch}}
\address{Institute for Theoretical Physics\\
		University of Bern\\
		Sidlerstrasse 5, CH-3012 Bern, Switzerland}

\author{C.J.~Isham\thanks{email: c.isham@ic.ac.uk}}
\address{Theoretical Physics Group,
		Blackett Laboratory\\
		Imperial College of Science, Technology \& Medicine\\
		South Kensington, London  SW7 2BZ, U.K.}

\date{October, 1995}
\maketitle

\begin{abstract}
The perennial formalism is applied to the real, massive Klein-Gordon
field on a globally-hyperbolic background space-time with compact
Cauchy hypersurfaces. The parametrized form of this system is taken
over from the accompanying paper. Two different algebras ${\cal
S}_{\text{can}}$ and ${\cal S}_{\text{loc}}$ of elementary
perennials are constructed. The elements of ${\cal S}_{\text{can}}$
correspond to the usual creation and annihilation operators for
particle modes of the quantum field theory, whereas those of ${\cal
S}_{\text{loc}}$ are the smeared fields. Both are shown to have the
structure of a Heisenberg algebra, and the corresponding Heisenberg
groups are described. Time evolution is constructed using
transversal surfaces and time shifts in the phase space. Important
roles are played by the transversal surfaces
associated with embeddings of the Cauchy hypersurface in the
space-time, and by the time shifts that are generated by
space-time isometries. The automorphisms of the algebras generated
by this particular type of time shift are calculated explicitly.

	The construction of the quantum theory using the perennial
formalism is shown to be equivalent to the Segal quantization of a
Weyl system if the time shift automorphisms of the algebra ${\cal
S}_{\text{can}}$ are used. In this way, the absence of any timelike
Killing vector field in the background space-time leads naturally to
the `problem of time' for quantum field theory on a background
space-time. Within the perennial formalism, this problem is formally
identical to the problem of time for any parametrized system,
including general relativity itself. Two existing strategies---the
`scattering' approach, and the `algebraic' approach---for dealing
with this problem in quantum field theory on a background space-time
are translated into the language of the perennial formalism in the
hope that this may give some insight into how the general problem can be
solved. The non-unitary time evolution typical of the Hawking effect
is shown to be due to global properties of the corresponding phase
space: specifically, the time shifts map a global transversal
surface to a non-global one. Thus, the existence of this effect is
closely related to the global time problem.
\end{abstract}
\pacs{}

\section{Introduction}
\label{Sec:intro}
In the canonical approach to quantum gravity, much emphasis is placed
on three particular issues: the conceptual problems that arise in
the interpretation of the theory, especially the problem of time;
the role of the space-time diffeomorphism group; and the
construction of non-perturbative quantization methods (for reviews,
see \cite{isham-rev1,isham-rev2,kuch-prehled}).  The Dirac method of
imposing operator constraints on the allowed state vectors (which,
in the case of gravity, leads to the Wheeler-DeWitt equation) has
become particularly popular because of its manifest relation to the
idea of invariance under the action of space-time diffeomorphisms.

	However, diffeomorphism invariance can also be secured by
adopting a method in which only `gauge-invariant' objects are
quantized; Dirac himself laid the foundations \cite{dirac} for a
powerful approach of this sort. A generalization of the Dirac method
to include all finite-dimensional, first-class parametrized systems
was presented by H\'{a}j\'{\i}\v{c}ek \cite{timelevels}. The
resulting theory, which combines the group-theoretic \cite{isham}
and algebraic \cite{blue-book} methods of quantization with that of
Dirac, is called the `perennial formalism', following the
notation introduced by Kucha\v{r} in his analysis
\cite{kuch-prehled} of the problem of time in canonical quantum
gravity. The key idea of this method is to find an algebra of
phase-space functions whose Poisson brackets with all the
first-class constraints vanish; such functions are therefore
constant on the phase-space orbits of the (function) group generated
by the constraints.  Furthermore, this algebra is required to be
large enough to generate all gauge-invariant functions in an
appropriate sense. Quantization of the system then consists in
finding irreducible self-adjoint representations of this algebra of
`physical observables' or, essentially equivalently, finding
irreducible, unitary representations of the associated `canonical'
group.

	In the present paper we develop the perennial formalism in the
context of a field system, namely a linear, massive scalar field
propagating on a fixed, globally-hyperbolic space-time with compact
Cauchy hypersurfaces.

	Various motivations lie behind such a study. To begin with,
infinite-dimensional systems are qualitatively different from
finite-dimensional ones, and it is an important---and mathematically
non-trivial---challenge to see how the perennial formalism can be
extended to this case. We shall show how the scalar field theory can
be rewritten in such a way as to become a simple example of a system
with perennials.  However, quantum field theory in a fixed
background has been much studied in the past using standard methods,
and hence it provides a useful model for exploring the perennial
formalism for an infinite-dimensional system.  As we shall see, new
problems {\em do\/} appear, the most important of which concerns the
choice of the operator representation for the group/algebra
generated by the perennials.

	In the standard approach to quantum field theory on a background
space-time, the normal way of addressing the problem of operator
representation makes extensive use of the classical time evolution
of the system. Thus the time evolution and associated Hamiltonian
are considerably more important for infinite-dimensional systems
than they are in the finite-dimensional case. Specifically, if there
is no timelike isometry group then the physical representation for
the quantum system is not determined and---at the same time---the
time evolution and Hamiltonian of the classical system are not
well-defined (by the perennial formalism). Such a lack of a
Hamiltonian was identified in \cite{timelevels} as the general
form taken by the problem of time within the perennial formalism.
However, studies of quantum field theory on a background space-time
have produced several possible strategies for dealing with this
problem, and one can hope that some of these ideas may be applicable
to other situations, especially if a common language---in our case,
that of the perennial formalism---has been developed.

	The perennial formalism enables us to reformulate the dynamics
of a scalar field in a background space-time in terms of properties
of the system's phase space. In this reformulation, an important
role is played by the idea of a {\em transversal surface\/}, defined
in general for a system with a gauge group as a submanifold in the phase
space that cuts orbits of the gauge group transversally; the
{\em domain\/} of a transversal surface is the set of points in the
constraint submanifold that can be joined to the transversal surface
by the orbits of the constraints, see \cite{timelevels}. Global
problems may arise: for example, there may not exist any global
transversal surface ({\em i.e.}, one whose domain is the whole
constraint surface); or there might be a symmetry transformation
that maps a global transversal surface onto a transversal surface
that is not global.  Indeed, we shall show that the latter situation
arises in the particular case of a space-time with a black hole.
This leads to an information loss in which the quantum evolution
associated with the symmetry sends pure states to mixed states (the
Hawking effect).  A situation with somewhat similar features can
occur if a system has no global transversal surfaces (the so-called
`global' time problem); in particular, an analogous non-unitarity is
exhibited by the quantum time evolution.  Toy models exhibiting such
behaviour were studied in \cite{patchI,patchII}. Thus, the ideas
developed in the present paper may be of use in finding a physical
interpretation in general situations in which the global time
problem arises.

	Another motivation for our present work is that our
understanding of quantum field theory on a fixed background might
itself profit from the use of the language of perennials.  For
example, the symmetries and transversal surfaces that have been
employed in the past in studies of such quantum field theories are
of a very special nature: namely, transformations of phase space that
are generated in a particular way by space-time transformations, and
surfaces associated with space-time hypersurfaces.  The perennial
formalism permits more general types of symmetry and transversal
surface, and suggests how these can be found. The question then is
if these symmetries can be utilized in the quantum theory, and---if
not---why not.

	The plan of the paper is as follows.  In section
\ref{Sec:phase}, we summarize the results of Ref. \cite{I+H} where
the dynamics and the symplectic geometry of the parametrized scalar
field was cast into the standard form of a first-class parametrized
system so that the perennial formalism can be applied. In section
\ref{Sec:perennials}, two different kinds of perennials are
constructed and each of them is shown to form an algebra of
elementary perennials, both of which are versions of the infinite
Heisenberg algebra.  We show that each isometry of the background
space-time defines a map of the phase space that is a symmetry.  We
calculate the action of these symmetries on the elementary
perennials and find that they define automorphisms of the algebras.
Then we show how the theory of time evolution as described in
\cite{timelevels} can be applied to the present case.

	In the final section we show how the group and algebraic
quantization that is performed as the next step reduces to the
familiar problem of finding a physically appropriate representation
of the Weyl group. We briefly summarise the relevant results of the
theory of Weyl systems in connection with quantum field theory on a
background space-time, and we show that our quantization method
leads to a quantum theory that is equivalent to the usual one.
Finally, we discuss the problem of time in a quantum field theory in
a fixed background.  We study a `scattering approach' to
quantization that makes use of an isolated symmetry that is defined
on only a small subset of the phase space.  In the context of the
perennial formalism, such a generalized symmetry is sufficient to
define a time evolution. We show that the resulting quantum
evolution is non-unitary if the symmetry does not preserve the
domains of the transversal surfaces involved in the construction,
and we apply the results to the Hawking effect. We also briefly
describe the `algebraic approach' to quantization in which the
states are defined as functionals on the algebras of perennials.

\section{Structure of the extended phase space}
\label{Sec:phase}
In this section, we shall summarize the results of the companion
work \cite{I+H} so that the present paper becomes self-contained.
For more details, one should consult \cite{I+H}.

	We work with a curved background space-time $({\cal M},g)$ and
assume that it is $C^{\infty}$ and globally hyperbolic; the Cauchy surface
$\Sigma$ is assumed to be compact. The real
scalar field $\phi$ satisfies the Klein-Gordon equation
\begin{equation}
   	|\det g|^{-1/2} \partial_\mu(|\det g|^{1/2}g^{\mu\nu}
	\partial_\nu\phi) + m^2\phi = 0.           		\label{K-G}
\end{equation}

	Consider a $C^{\infty}$ embedding $X:\Sigma\rightarrow {\cal M}$
that is spacelike with respect to the metric $g$. Let $\cal E$
denote the space of all such embeddings. Each embedding $X$
determines a positive-definite metric $\gamma_X$ on $\Sigma$ and a
unit normal vector field $n_X$ to $X(\Sigma)$ in $\cal M$. The
embedding $X$ also defines a Cauchy datum for the field $\phi$ along
the hypersurface $X(\Sigma)$. This is a pair $(\varphi,\pi)$ of
fields on $\Sigma$, where the scalar $\varphi$ and the density (of
weight $w=1$) $\pi$ are defined by
\begin{eqnarray}
    \varphi(x)	& := & \phi(X(x)),
\label{Def:varphi}\\
    \pi(x)      & := & (\det\gamma)^{1/2}(X(x))\, n^{\mu}(X(x))\,
					\partial_{\mu}\phi(X(x)).
\label{Def:pi}
\end{eqnarray}
The space of all $C^{\infty}$ Cauchy data will be denoted by
$\Gamma_{\phi}$---a linear space that can be equipped with a
Sobolev structure (for details, see \cite{I+H}). The dynamical
equation (\ref{K-G}) defines a mapping between Cauchy data
corresponding to different embeddings. Specifically, let $X$ and $X'$
be two arbitrary spacelike embeddings and let
$(\varphi,\pi)\in\Gamma_{\phi}$. Then there is a unique solution
$\phi$ of equation (\ref{K-G}) whose Cauchy datum at $X(\Sigma)$ is
$(\varphi,\pi)$, and this induces a well-defined Cauchy datum
$(\varphi',\pi')$ on $X'(\Sigma)$. Thus we get a map $(\varphi,\pi)
\rightarrow (\varphi',\pi')$, which we denote by
$\rho_{X\,X'}$.  One can show that $\rho_{X\,X'}$ is an
automorphism of the Sobolev space $\Gamma_{\phi}$.

	The space $\Gamma_{\phi}$ is the phase space of the
(non-constrained) scalar field $\phi$ on the curved background
$({\cal M},g)$. If we extend this space by adding all spacelike
embeddings $X$ and their conjugate momenta $P$, and if we impose
suitable constraints, we obtain a constrained system that is
dynamically equivalent to the original one. The points of the
resulting extended phase space $\Gamma$ are collections of fields,
$x\mapsto(\varphi(x),\pi(x),X(x),P(x))$ on $\Sigma$. These fields
can be characterised by their transformation properties with respect
to a pair of local charts
\begin{equation}
    (U,h){\text{ of }}\Sigma{\text{ and }}(\bar{V},\bar{h})
		{\text{ of }} {\cal M} 			\label{charts}
\end{equation}
(where $X(U)\cap\bar{V}\neq\emptyset$). In particular, $P_{\mu}(x)$
is a covector with respect to the transformation of
$(\bar{V},\bar{h})$ on $\cal M$ and a quadruple of scalar densities
with respect to the transformation of $(U,h)$ on $\Sigma$.
Quantities of this type were called `$e$-tensor densities' by
Kucha\v{r} \cite{hyperspace}.

	The phase space $\Gamma$ can be given a structure of an
infinite-dimensional differentiable manifold with tangent and
cotangent vectors described as follows. Consider a curve
$\lambda\rightarrow
(\varphi_{\lambda},\pi_{\lambda},X_{\lambda},P_{\lambda})$ whose
tangent vector components
$(\Phi,\Pi,V,W)\equiv
(\dot{\varphi}_{\lambda},\dot{\pi}_{\lambda},\dot{X}_{\lambda},
\dot{P}_{\lambda})$
can be calculated by differentiating with respect to $\lambda$ the
coordinate representatives associated with the pair of charts
(\ref{charts}). The fields $\Phi(x)$, $\Pi(x)$, $V(x)$ and $W(x)$
are again characterized by their transformation properties: the
first three are $e$-tensors, but the fourth transforms in a more
complicated way (see \cite{I+H}). The space
$T_{(\varphi,\pi,X,P)}\Gamma$ of all such vectors can be given the
$C^{\infty}$ structure of a Fr\'{e}chet space.  A cotangent vector
at a point $(\varphi,\pi,X,P)$ of $\Gamma$ will be a quadruple
$(A_{\varphi},A_{\pi},A,B)$ of fields on $\Sigma$ such that the
pairing
\begin{equation}
 \langle(A_{\varphi},A_{\pi},A,B),(\Phi,\Pi,V,W)\rangle :=
     \int_\Sigma d^3 x\,(A_{\varphi}\Phi+A_{\pi}\Pi+A_\mu V^{\mu}+
		     B^\mu W_{\mu}).
\end{equation}
with vectors from $T_{(\varphi,\pi,X,P)}\Gamma$ gives a coordinate
independent number. This requirement determines the transformation
properties of the fields $(A_{\varphi},A_{\pi},A,B)$.

One can show that $(A_{\pi},-A_{\varphi},B,-A)$ transforms as a
tangent vector. Thus, there is a map
$J:T_{(\varphi,\pi,X,P)}^*\Gamma\rightarrow
T_{(\varphi,\pi,X,P)}\Gamma$ given by
$J(A_{\varphi},A_{\pi},A_X,A_P) := (A_{\pi},-A_{\varphi},A_P,-A_X)$.
The $C^{\infty}$ structure of the space
$T_{(\varphi,\pi,X,P)}^*\Gamma$ can be chosen in such a way that $J$
is an isomorphism.  Using this isomorphism, a symplectic structure
$\Omega$ on $\Gamma$ can be defined as follows. If $v_1$ and $v_2$
are two vectors in $T_{(\varphi,\pi,X,P)}\Gamma$ then
\begin{equation}
     \Omega(v_1,v_2) := -\langle J^{-1}v_1, v_2\rangle	\label{Def:Omega}.
\end{equation}
It follows at once that (i) $\Omega(v_1,v_2 ) = -\Omega(v_2,v_1)$; (ii)
$\Omega$ is weakly non-degenerate (see \cite{C+M}); and (iii) $\Omega$ is not
only closed but also exact.

	As $\Omega$ is only a weak symplectic form, not every
differentiable function on $\Gamma$ will have an associated
Hamiltonian vector field.  The class of functions that do can be
characterized as follows.  If $F:\Gamma\rightarrow\mathR$, we say
that $F$ has a {\em gradient\/} if the following two conditions are
satisfied:
\begin{enumerate}
     \item the Fr\'{e}chet derivative, ${\text{D}}F|_{(\varphi,\pi,X,P)}:
			T_{(\varphi,\pi,X,P)}\Gamma\rightarrow\mathR$ is a
			bounded linear map;
     \item there exists ${\text{grad}}\,F\in
		T_{(\varphi,\pi,X,P)}^*\Gamma$ such that
          $\langle{\text{grad}}\,F,v\rangle
		 = {\text{D}}F|_{(\varphi,\pi,X,P)}(v)$ for all
		$v\in T_{(\varphi,\pi,X,P)}\Gamma$.
\end{enumerate}
The quantity ${\text{grad}} F$ is calculated from ${\text{D}}F$ as usual by
integration by parts (if $F$ contains derivatives). The `components' of this
gradient will be denoted by the collection of functions
			$({\text{grad}}_{\varphi}F, {\text{grad}}_{\pi}F,
	          	({\text{grad}}_X F)_\mu,({\text{grad}}_P F)^\nu)$.

	For a differentiable function with a gradient, we can define the
associated `Hamiltonian vector field'.  Specifically, if $F$ is such
a function, then $\xi_F\in T_{(\varphi,\pi,X,P)}\Gamma$ is defined
by the relation $\langle{\text{grad}}\,F, v\rangle =
\Omega(v,\xi_F),{\text{ for all }} v\in
T_{(\varphi,\pi,X,P)}\Gamma$. Hence, because of Eq.\
(\ref{Def:Omega}), we see that $\langle{\text{grad}}\,F, v\rangle =
\langle J^{-1}\xi_F, v\rangle$ for all $v$, and so $\xi_F =
J({\text{grad}}\,F)$. Finally, the Poisson bracket of a pair of
differentiable functions $F$ and $G$ is defined as $\{F,G\} :=
-\Omega(\xi_F,\xi_G)$, and we see immediately that
\begin{equation}
     \{F,G\} = \langle{\text{grad}}\,F,\xi_G\rangle. \label{poisson}
\end{equation}
This Poisson bracket is antisymmetric and, since $\Omega$ is closed, it
satisfies the Jacobi identity.

	The constraints that must be imposed on the extended phase space
$\Gamma$ are
\begin{eqnarray}
  {\cal H}_{\mu} 		&=& P_{\mu}+{\cal H}_{\mu}^{\phi}, \\
								\label{cst1}
  {\cal H}_{\mu}^{\phi} & = & -{\cal H}_{\bot}^{\phi}\,n_{\mu} +
				{\cal H}_k^{\phi}\,X_{\mu}^k,	\label{cst2}
\end{eqnarray}
where $ {\cal H}_{\bot}^{\phi} = \frac{1}{2}(\det\gamma)^{1/2}
\left(\frac{\pi^2}{\det\gamma} + \gamma^{kl}\,\varphi,_k\varphi,_l
 + m^2\,\varphi^2 \right)$, and ${\cal H}_k^{\phi} =
\pi\,\varphi,_{k}$.  One can show that the constraint set
$\tilde\Gamma$ defined by ${\cal H}_\mu=0$ is a smooth submanifold
in a neighbourhood of any of its points.  Moreover, $\tilde{\Gamma}
= \tilde{C}(\Gamma_{\phi} \times {\cal E})$, where
$\tilde{C}:\Gamma_{\phi}\times{\cal E}\rightarrow\Gamma$ is defined
by $\tilde{C}(\varphi,\pi,X):= (\varphi,\pi,X,-{\cal H}^{\phi}
(\varphi,\pi,X))$. We shall often refer to $\tilde{\Gamma}$ as
$\Gamma_{\phi}\times{\cal E}$.

	If we smear ${\cal H}_{\mu}$ to get ${\cal H}_N :=
\int_{\Sigma}d^3x\, N^{\mu}(x){\cal H}_{\mu}(x)$, where $N\in
T_X{\cal E}$ ({\rm i.e.}, $N(x)\in T_{X(x)}{\cal M}$), we obtain a
differentiable function with a gradient. Such a function defines a
Hamiltonian vector field, and one can proceed in complete analogy
with the theory of finite-dimensional systems.  For example, let
$\phi$ satisfy the Klein-Gordon equation (\ref{K-G}) on $\cal M$.
Then for each curve $\lambda\mapsto X_{\lambda}$ with the tangent
vector field $N$ on $\cal E$, the initial data $(\varphi_{\lambda},
\pi_{\lambda})$ for $\phi$ on $X_\lambda(\Sigma)$ satisfy the
evolution equation
\begin{equation}
	(\dot{\varphi},\dot{\pi},\dot{X},\dot{P}) =
		J({\text{grad}}{\cal H}_N).
\label{evoleq}
\end{equation}
Thus the Hamiltonian vector fields $J({\text{grad}}{\cal H}_N)$ of
the functions ${\cal H}_N$ are tangential to $\tilde{\Gamma}$, and
the system is first class according to the definition given in
\cite{timelevels}. The pull-back of the vector field (\ref{evoleq})
to $\Gamma_{\phi}\times\cal E$ is given by
\begin{eqnarray}
     \dot{\varphi} & = & {\text{grad}}_{\pi}{\cal H}_N , \label{evoleqFi} \\
     \dot{\pi} & = & -{\text{grad}}_{\varphi}{\cal H}_N, \label{evoleqPi} \\
     \dot{X} & = & N.
\end{eqnarray}

	Let us denote the space of longitudinal vectors at
$(\varphi,\pi,X) \in \Gamma_{\phi}\times \cal E$ by
$\Xi_{(\varphi,\pi,X)}$, {\em i.e.},
\begin{equation}
     \Xi_{(\varphi,\pi,X)} := \{(\Phi,\Pi,V)\in T_{(\varphi,\pi,X)}
		\tilde{\Gamma}\mid 	\Phi = {\text{grad}}_{\pi}{\cal H}_N,
				\Pi = -{\text{grad}}_{\varphi}{\cal H}_N,
					V = N\in T_X{\cal E}\}. \label{Xi}
\end{equation}
The space $\Xi_{(\varphi,\pi,X)}$ is a closed subspace of
$T_{(\varphi,\pi,X)}\tilde{\Gamma}$. Moreover, there is a
submanifold of $\tilde{\Gamma}$ whose tangent space coincides with
$\Xi_{(\varphi,\pi,X)}$ at each point of the submanifold. Let us
denote the maximal submanifold of this kind passing through a point
$(\varphi,\pi,X)\in\tilde{\Gamma}$ by $\gamma_{(\varphi,\pi,X)}$.
This subset $\gamma_{(\varphi,\pi,X)}$ is called a `$c$-orbit through
$(\varphi,\pi,X)$', in complete analogy with the situation for a
finite-dimensional system.

	Finally, the pull-back $\tilde{\Omega}$ of the form $\Omega$ to
$\tilde{\Gamma}$ is given by the formula
\begin{eqnarray}
\lefteqn{\tilde{\Omega}((\Phi_1,\Pi_1,V_1 ), (\Phi_2,\Pi_2,V_2)) =}
								\nonumber \\
&&\int_\Sigma d^3x\,[(\Pi_1+{\text{grad}}_{\varphi}{\cal H}_{V_1})
(\Phi_2 - {\text{grad}}_{\pi}{\cal H}_{V_2}) - (\Pi_2 +
{\text{grad}}_{\varphi}{\cal H}_{V_2}) (\Phi_1 -
{\text{grad}}_{\pi}{\cal H}_{V_1})]. \label{pulback}
\end{eqnarray}
Clearly, $\tilde{\Omega}$ is a presymplectic form and
$\Xi_{(\varphi,\pi,X)}$ is its singular subspace.

\section{Perennials, symmetries and time evolution}
\label{Sec:perennials}
In section \ref{Sec:phase} we showed that the geometrical structure
of the phase space, constraint submanifold and the $c$-orbits of our
infinite-dimensional system are all analogous to those of the
corresponding objects in a finite-dimensional system as studied, for
example, in \cite{timelevels}. The application of the perennial
formalism is now straightforward.

	A crucial role in the perennial formalism is played by
quantities that are reparametrization and gauge invariant. In
particular, a {\em perennial\/} is defined as a function $o:\Gamma
\rightarrow\mathR$ that is constant along the $c$-orbits; or, equivalently,
\begin{equation}
     \{o,H_N\}|_{\tilde{\Gamma}} = 0, {\text{ for all }} N.
\end{equation}

	In most physical applications of a field theory, one deals with
a restricted class of functions on the phase space---the so-called
`local functionals'. Each local functional has the form
$\int_{\Sigma}{\text d}^3 x\,F(x)$, where the value of $F(x)$ at
$x\in\Sigma$ is a polynomial function of values of the fields and
their $x$-derivatives taken at the same point. It is easy to show
that (i) all local functionals possess gradients; (ii) the Poisson
bracket of two local functionals is again a local functional, so
that multiple Poisson brackets are well-defined; and (iii) they
satisfy the Jacobi identity. Since the smeared constraint ${\cal
H}_N$ is itself a local functional it follows that the set of all
local functionals that are perennials forms a Poisson algebra ${\cal
P}_{\text {lf}}$.

	Let us now construct a particular class of local functional
perennials for the case of scalar field theory on a background
space-time.  The idea is to associate a perennial $o_\phi$ with each
maximal solution $\phi$ of the classical field equations.
Specifically, let $(\xi,\eta,X,P)$ be an arbitrary point of $\Gamma$
and let $(\varphi,\pi )$ be the Cauchy datum of $\phi$ at $X$. Then
$o_\phi$ is defined by
\begin{equation}
     o_\phi(\xi,\eta,X,P) :=
	\int_\Sigma d^3x\,(\varphi\eta-\xi\pi). 		\label{ophi}
\end{equation}
The main task is to show that $o_\phi$ is constant along $c$-orbits.

	Let $\gamma_{(\xi ,\eta ,X)}$ be a $c$-orbit and let $\psi$ be the
associated maximal classical solution, {\em i.e.}, $(\xi,\eta)$ is
the Cauchy datum of $\psi$ on $X(\Sigma)$. Let
$(\xi',\eta',X',P')\in\gamma_{(\xi,\eta,X)}$ be an arbitrary point on
the $c$-orbit. Then
\begin{equation}
     o_\phi(\xi',\eta',X',P') =
	\int_\Sigma d^3 x\,(\varphi'\eta' -\xi'\pi'),
\end{equation}
where $(\varphi',\pi')$ and $(\xi', \eta')$ are the Cauchy data of $\phi$
and $\psi$ respectively at $X'$. Using Eqs.\ (\ref{Def:varphi}) and
(\ref{Def:pi}), we can rewrite this as
\begin{equation}
     o_\phi(\xi',\eta',X',P') = \int_{\Sigma}d^3 x\,(\det \gamma')^{1/2}
     n^{'\mu}(\phi\psi_{,\mu} - \psi\phi_{,\mu})_{X'(\Sigma)}
\end{equation}
where $\gamma'$ is the induced metric on $X'(\Sigma )$, and
$n^{'\mu}$ is the unit normal vector to $X'(\Sigma )$. However,
the integral on the right hand side is just the familiar
`Klein-Gordon inner product' $(\phi,\psi )_{KG}$ of the two solutions
$\phi$ and $\psi$, and this is well-known to be independent of the
Cauchy surface $X'(\Sigma)$.  Hence $o_\phi$ is constant, as
claimed.

	The perennial $o_\phi$ has the following properties.
\begin{enumerate}
\item Suppose that $o_\phi$ is constant along the whole of
     the constraint set $\tilde{\Gamma}$. Then $\phi$
     must have the same Klein-Gordon product with any other
	solution, which is only possible if $\phi = 0$.

\item Let $\phi$ and $\psi$ be two maximal solutions with corresponding
	perennials  $o_\phi$ and $o_\psi$. Then the definition
     (\ref{ophi}) implies immediately that
     \begin{eqnarray}
          o_\phi + o_{\psi} & = & o_{\phi + \psi}, \label{S1} \\
          r o_\phi & = & o_{r\phi} {\text{ for all }}r\in\mathR, \label{S2}
     \end{eqnarray}
     where the linearity of the field equation (\ref{K-G}) guarantees
	that the solutions $\phi+\psi$ and $r\phi$ are again maximal.

\item The Poisson bracket of $o_\phi$ and $o_{\psi}$ can be obtained from
     Eq.\ (\ref{poisson}). To calculate it we need the gradients, and
     using Eqs.\ (\ref{evoleqFi}) and (\ref{evoleqPi}) we obtain
     \begin{eqnarray}
          \lefteqn{\langle{\text{grad}}\,o_\phi|_{(\xi,\eta,X,P)}\,,\,
		 				(\Phi,\Pi,V,W)\rangle =} \nonumber \\
          & & \int_\Sigma d^3 x\,[\varphi\Pi - \pi\Phi +
          \xi\,{\text{grad}}_{\varphi}H_V|_{(\varphi,\pi,X,P)} +
          \eta\,{\text{grad}}_{\pi}H_V|_{(\varphi,\pi,X,P)}].
     \end{eqnarray}
     Thus,
     \begin{eqnarray*}
          {\text{grad}}_{\varphi}o_\phi|_{(\xi ,\eta ,X,P)} & = & - \pi , \\
          {\text{grad}}_{\pi}o_\phi|_{(\xi,\eta,X,P)}&=& \varphi , \\
          {\text{grad}}_P o_\phi|_{(\xi,\eta,X,P)} & = & 0,
     \end{eqnarray*}
     where $(\varphi,\pi )$ is the Cauchy datum of $\phi$ at $X$. It
     follows that
     \begin{equation}
          J({\text{grad}}\,o_\phi) = (\varphi,\pi,0,A), \label{Ido}
     \end{equation}
     where $A_{\mu}$ are functions of $\xi$, $\eta$, $X$, $P$ and
     $\phi$. Then
     \begin{equation}
          \{o_\phi,o_{\psi}\}(\xi,\eta,X,P) =
          \int_{\Sigma}d^3 x\,(\det\gamma)^{1/2}n^{\mu}(\phi\psi_{,\mu}
          -\psi\phi_{,\mu})_{X(\Sigma )}=(\phi,\psi)_{KG}, 	\label{S3}
     \end{equation}
     where $\gamma$ and $n^{\mu}$ are the induced metric and unit normal
     vector at $X(\Sigma )$. Thus the Poisson bracket is independent of
     $(\xi,\eta ,X,P)$ and is hence a constant real function on the phase
	space $\Gamma$.

\item Let $\phi$ and $\psi$ be two different maximal solutions
	 representing two different
     orbits $\gamma_\phi$ and $\gamma_{\psi}$. Then there is a third
     solution, $\chi$, such that $(\chi,\phi-\psi )_{KG} \neq 0$,
     and hence $o_{\chi}$ has different values at $\gamma_\phi$ and
     $\gamma_{\psi}$.
\end{enumerate}

	Let ${\cal S}_{\phi}$ denote the set of perennials of the form
$o_{\phi}$ where $\phi$ runs over the set of all $C^{\infty}$
solutions to the field equations. Then, because of the second
property above, ${\cal S}_\phi$ is a linear space. Let ${\cal R}
\cong\mathR$ denote the set of the constant real functions on
$\Gamma$, and consider the linear space ${\cal S}_{\text{can}} :=
{\cal S}_{\phi} \oplus {\cal R}$. This space ${\cal S}_{\text{can}}$
is closed with respect to Poisson bracket operations because of the
third property above.  Moreover, ${\cal S}_{\text{can}}\subset {\cal
P}_{\text{lf}}$. Thus ${\cal S}_{\text{can}}$ is a Lie subalgebra
of ${\cal P}_{\text{lf}}$. According to the fourth property above,
it separates the $c$-orbits in $\tilde{\Gamma}$. It follows that ${\cal
S}_{\text{can}}$ can play the role of an `algebra of elementary
perennials' for our system---the basic ingredient in the `algebraic
method of quantization' in which the quantum theory is associated
with a self-adjoint representation of ${\cal S}_{\text{can}}$ on a
Hilbert space (see \cite{timelevels}). The relations (\ref{S1}),
(\ref{S2}) and (\ref{S3}) imply that ${\cal S}_{\text{can}}$ is
an infinite-dimensional Heisenberg algebra. Specifically, as
a linear space it is a direct sum of $\cal R$ and the space ${\cal
S}_{\phi}$ that is equipped with the (weakly) non-degenerate
skew-symmetric form $(\cdot,\cdot)_{KG}$; the Lie
bracket is then defined by
\begin{equation}
  [(\phi_1,r_1 ),(\phi_2 ,r_2)] := (0,(\phi_1,\phi_2 )_{KG}).
\end{equation}
The corresponding Lie group is the so-called `Heisenberg group' ${\cal
G}_{\text{can}}$ defined on ${\cal S}_{\phi}\times\mathR$ by the group law
\begin{equation}
  (\phi_1,r_1)\cdot(\phi_2,r_2) :=
	(\phi_1 + \phi_2, r_1 + r_2 + \frac{1}{2}(\phi_1,\phi_2 )_{KG}).
\end{equation}

	The action of ${\cal G}_{\text{can}}$ on $\Gamma$ can be deduced
from the action of its generators $o_{\phi}$ given by equation
(\ref{ophi}): namely, the point $(\xi,\eta,X,P)$ maps to $(\xi +
\varphi,\eta +\pi,X,P')$, where $(\varphi,\pi)$ is the initial datum
of $\phi$ at $X(\Sigma )$ and $P'$ is a function of $\xi,\eta,X,P$,
and $\phi$ such that $(\xi + \varphi, \eta +\pi,X,P')
\in{\tilde{\Gamma}}$ if $(\xi,\eta,X,P)\in{\tilde{\Gamma}}$. The
action is not faithful since the subgroup $(0,\mathR )$ acts
trivially, and hence the group ${\cal G}_{\text{can}}$ is a central
extension of a group of symmetries on $\Gamma$ (see \cite{isham}).
Thus ${\cal G}_{\text{can}}$ satisfies all the conditions for a
so-called `first-class canonical group' whose irreducible, unitary
representations can be associated with a quantization of the system
\cite{isham}.

	There is an alternative choice for the algebra of elementary
perennials in which the perennials are associated with `smeared
fields'; as such, they form the basis for a different (but
ultimately equivalent) quantization of the scalar field. The
construction goes as follows. Let $D({\cal M})$ be the space of
$C^{\infty}$ test functions with compact support on the space-time
${\cal M}$, let $f\in D({\cal M})$ and
$(\varphi,\pi,X,P)\in\Gamma$. Then there is a unique maximal
classical solution $\phi$ with the Cauchy datum $(\varphi,\pi)$ at
$X(\Sigma)$, and we define the perennial $\kappa_f:
\Gamma\rightarrow\mathR$ by the equation
\begin{equation}
  \kappa_f(\varphi,\pi,X,P) :=
\int_{\cal M}{\text d}^4 y\,|\det g|^{1/2}\phi f. \label{smear}
\end{equation}
Note that $\kappa_f$ does not depend on $P$, and it is a perennial
because the same classical solution leads to the same value of
$\kappa_f$. Let us list some important properties of this type of
perennial.
\begin{enumerate}
\item
	Clearly, $\kappa_f$ can be constant along ${\tilde{\Gamma}}$ only
	if $f=0$, and then $\kappa_f=0$.
\item
	Let $f$ and $f'$ be two elements of $D({\cal M})$ with
	corresponding perennials $\kappa_f$ and $\kappa_{f'}$. Then the
	definition (\ref{smear}) implies immediately that
	\begin{eqnarray*}
	  \kappa_f + \kappa_{f'} & = &\kappa_{f + f'}, \\
	  r\kappa_f & = & \kappa_{rf}{\text{ for all }}r\in\mathR.
	\end{eqnarray*}
\item
	{We can find an explicit expression for $\kappa_f$ if we use the
	Cauchy propagator $G(x,y)$ for the equation (\ref{K-G})
	($G(x,y)$ is sometimes known as the `Pauli-Jordan function'). The
	existence and uniqueness of such a propagator for
	space-times of the type with which we are dealing was shown by
	Choquet-Bruhat \cite{C-B}. The basic properties of the Cauchy
	propagator (for example, see \cite{dewitt}) are (i)
	$G(x,y) = -G^r (x,y) + G^a (x,y)$,
	where $G^r$ and $G^a$ are respectively the retarded and advanced
	propagators; (ii) $G(x,y)$ is real and skew-symmetric in $x$ and $y$;
	and (iii) $G(x,y)$ satisfies the identity
	\begin{equation}
	  \phi (x) = (G(x,\cdot),\phi(\cdot))_{KG}, 	\label{propag}
	\end{equation}
	where $\phi (x)$ is any $C^{\infty}$ solution to equation
	(\ref{K-G}). From Eq.\ (\ref{propag}), it follows immediately that
	\begin{equation}
	  (G(x,\cdot),G(\cdot,y))_{KG} = G(x,y).		\label{compos}
	\end{equation}

	If the Klein-Gordon product on the right hand side
	of Eq.\ (\ref{propag}) is written
	out along $\Sigma$, we obtain an expression for the solution $\phi$
	at any point $y\in{\cal M}$ in terms of its Cauchy data at
	$X(\Sigma )$. Substituting for $\phi(y)$ in Eq.\ (\ref{smear}) from
	Eq.\ (\ref{propag}) then gives the desired formula:
	\begin{equation}
	  \kappa_f = -\int_{\Sigma}{\text d}^3 x (\det \gamma )^{1/2}n^{\mu}
		\left.\frac{\partial G(f,y)}{\partial y^{\mu}}\right|_{X(x)}
		\varphi(x) +
			\int_{\Sigma}{\text d}^3 x\,G(f,X(x))\pi (x),
	\label{Phif}
	\end{equation}
	where
	\begin{equation}
	  G(f,x) := \int_{\cal M}{\text d}^4 y\,|\det g|^{1/2}f(y)G(y,x).
	\end{equation}
	Thus, $\kappa_f$ belongs to the class of local functionals.
	}
\item
	We have the relation $\{\kappa_f,\kappa_{f'}\} = -G(f,f')$
	whose derivation is simple: read off the gradient
	of $\kappa_f$ from the formula (\ref{Phif}), insert it in the
	equation (\ref{poisson}), and use the identity (\ref{compos}).
\end{enumerate}

	The smeared perennials generate a Lie algebra, which we denote
by ${\cal S}_{\text{loc}}$. Properties 2 and 4 above imply that
${\cal S}_{\text{loc}}$ is a Heisenberg algebra on $D({\cal
M})\times\mathR$ with the skew-symmetric form $-G(f,f')$.
The corresponding Heisenberg group ${\cal G}_{\text{loc}}$ can
be used as a first-class canonical group for the system.

	The next important step is to consider the role played by
symmetries, where---in complete analogy with the finite-dimensional
case (see \cite{timelevels})---a symmetry is defined as a symplectic
diffeomorphism of $\Gamma$ that preserves the constraint surface
$\tilde{\Gamma}$. In particular, it can be shown that each symmetry
maps $c$-orbits onto $c$-orbits.

	We shall describe a particular class of symmetries that play an
important role in the study of quantum field theory on a curved
space-time.  Any isometry $\vartheta:{\cal M}\rightarrow{\cal M}$
defines a map $\theta:\Gamma\rightarrow\Gamma$ as follows. Let
$(\varphi,\pi,X,P)\in \Gamma$ be arbitrary and set $X' :=
\vartheta\circ X$. Since $\vartheta$ is an isometry, the embedding
$X'$ is spacelike, and hence a Cauchy surface for $\cal M$. The
fields $\varphi$, $\pi$ and $P$ are $e$-tensor densities at
$X\in\cal E$, and so can be considered as $\cal M$-tensors  at
points in $X(\Sigma)$ (note that $\varphi(x)$ and $\pi(x)$ are
scalars, and $P(x)$ is a covector). Set $(\varphi',\pi',P') :=
(\vartheta^{\ast -1}\varphi,\vartheta^{\ast -1}
\pi ,\vartheta^{\ast -1}P)$, where $\vartheta^{\ast}$ is the usual
pull-back of differential forms on $\cal M$.  Finally, define
$\theta (\varphi,\pi,X,P) := (\varphi',\pi',X',P')$.

	A simple way of showing that $\theta$ is a symmetry is to use
the $\vartheta$-shifted chart. Each pair of local charts of the type
$(U,h),(\bar{V},\bar{h})$ in Eq.\ (\ref{charts}) can be `shifted' by
$\vartheta$ to become the chart $(U,h)$ on $\Sigma$ and the chart
$(\vartheta(\bar{V}),\bar{h}\circ\vartheta^{-1})$ on $\cal M$. The
functions that represent $(\varphi',\pi',X',P')$ in the shifted
charts coincide numerically with those that represent
$(\varphi,\pi,X,P)$ in the original charts. Moreover, the metric
$g'_{\mu\nu}$ in $\vartheta(\bar{V})$ coincides with $g_{\mu\nu}$ in
$\vartheta(\bar{V})$.  Thus, if $(\varphi,\pi,X,P)$ satisfies the
constraints, then $(\varphi',\pi',X',P')$ will also do so.
Furthermore, any curve
$\lambda\mapsto(\varphi_{\lambda},\pi_{\lambda},
X_{\lambda},P_{\lambda})$ on $\Gamma$ defines a curve
$\lambda\mapsto \theta
(\varphi_{\lambda},\pi_{\lambda},X_{\lambda},P_{\lambda})$ that has
the same form in the respective coordinate systems. Thus the tangent
vectors of these two curves must have the same components. It
follows that $\theta$ is differentiable and---moreover---symplectic
since the values of the symplectic form at $(\varphi,\pi,X,P)$ and
at $(\varphi',\pi',X',P')$ must coincide numerically in the
respective coordinate systems. Hence $\theta$ is a symmetry.

	Let us list some of the important properties of $\theta$.
\begin{enumerate}
\item The restriction $\tilde\theta:\Gamma_\phi\times{\cal E}\rightarrow
      \Gamma_\phi\times {\cal E}$ of $\theta$ to $\tilde\Gamma$,
     is given by
     \begin{equation}
          \tilde{\theta}(\varphi,\pi,X) = (\varphi,\pi,\vartheta\circ X).
     \end{equation}
     This follows immediately from the idea of a shifted chart and the
	fact that $\varphi$ and $\pi$ are scalar fields. Thus we obtain the
	same Cauchy datum at the shifted Cauchy surface.

\item Let $\phi$ be a global solution of the field equation,
	with a Cauchy datum in the Sobolev space
     $\Gamma_\phi$. Then $\phi\circ \vartheta^{-1}$ is again such a
	solution since $\vartheta$ is an isometry. If $\phi$ has the Cauchy
	datum $(\varphi,\pi)$ at $X$, then $\phi\circ \vartheta^{-1}$ has the
	datum $(\varphi,\pi)$ at $\vartheta\circ X$. It follows that the map
	$\rho_{XX'}$ defined in theorem 1 satisfies
	$\rho_{\theta\circ X\,\theta\circ X'}(\varphi,\pi) =
			\rho_{XX'}(\varphi,\pi)$.

\item The definition of $\theta$ implies immediately that it maps
	 $c$-orbits onto $c$-orbits. In particular, if $\gamma_\phi$ is an
	orbit corresponding to a maximal solution $\phi$, then
	$\theta(\gamma_\phi) = \gamma_{\phi\circ\vartheta^{-1}}$.

\item Let $o_\phi$ be a perennial in ${\cal S}_{\text{can}}$. The
     $\theta$-shifted perennial $s_{\vartheta}(o_\phi)$ was defined
     in \cite{timelevels} as $s_{\vartheta}(o_\phi) :=
	 o_\phi\circ\theta^{-1}$. Then we have the relation
     \begin{equation}
          s_{\vartheta}(o_\phi) = o_{\phi\circ\vartheta^{-1}}.
          \label{sper}
     \end{equation}
     Indeed, $o_\phi(\theta^{-1}(\xi,\eta,X,P)) =
	 o_\phi(\xi,\eta,X',P')=\int_\Sigma d^3 x\,(\varphi'\eta-\pi'\xi)$
     where $(\varphi',\pi')$ is the Cauchy datum of $\phi$ at $X' =
     \vartheta^{-1}\circ X$. However, $(\varphi',\pi' )$ is also the
     Cauchy datum of $\phi\circ\vartheta^{-1}$ at $X$. Thus,
     $o_\phi(\theta^{-1}(\xi ,\eta ,X,P)) =
			o_{\phi\circ\vartheta^{-1}}(\xi,\eta ,X,P))$,
     and this is equivalent to Eq.\ (\ref{sper}). A straightforward
	calculation gives the $\theta$-shift for ${\cal S}_{\text{loc}}$ as
	$s_{\vartheta}(\kappa_f) = \kappa_{f\circ\vartheta}$.

\end{enumerate}

	The relation in Eq.\ (\ref{sper}) implies that $s_{\vartheta}$
is an automorphism of the algebra ${\cal S}_{\text{can}}$ since the
map $\phi\mapsto\phi\circ \vartheta^{-1}$ is a linear transformation
of solutions that preserves the Klein-Gordon product, and the
constant functions on $\Gamma$ are left invariant by
$s_{\vartheta}$. Similarly, $s_{\vartheta}$ is an automorphism of
the algebra ${\cal S}_{\text{loc}}$ because the map $f\rightarrow
f\circ\vartheta $ is linear and preserves the quadratic form
$G(\cdot,\cdot )$. In fact, $s_{\vartheta}$ induces a
transformation of perennials from ${\cal S}_{\text{can}}$ that is
directly related to the Bogoliubov transformations that arise in the
study of quantum field theory on a curved background.  Indeed, if we
choose a complex orthonormal basis $\{\phi_m,\phi^*_m\}$ for the
space of solutions (for example, see \cite{dewitt}), then the coefficients
$\{a_m ,a^*_m\}$ of the expansion of any solution $\phi$ in terms of
this basis have the form of our perennials: namely $a_m =
(\phi^*_m,\phi)_{KG}$. The transformation $s_{\vartheta}$ of
perennials thus defines new coefficients $a_m$, and the expansion of
these in terms of the old ones is what is normally called a
`Bogoliubov transformation'.

	Let us observe that the perennial formalism allows a more
general type of symmetry that is not necessarily associated with
transformations of space-time. For example, the group
${\cal G}_{\text{can}}$ is a group of such symmetries. This raises the
interesting question of whether other symmetries that are not
associated with space-time transformations can be found,
and---if so---if they can be helpful in the study of quantum field theory
on a curved background.

	Finally, let us construct the time evolution of the system. In
the finite-dimensional case, such a construction is based on a
transversal surface $\Gamma_0$ and a one-dimensional symmetry group
$\{h(t)\}$ that moves $\Gamma_0$ (see \cite{timelevels}). A
transversal surface $\Gamma_0$ is defined to be a smooth submanifold
of $\tilde{\Gamma}$ that (i) intersects each $c$-orbit $\gamma$ in
at most one point $p=\Gamma_0\cap\gamma$; and (ii) has the property
that each such intersection is transversal, {\em i.e.},
$T_p\Gamma_0\cap T_p\gamma =\{0\}$, where $0$ is the zero vector. A
transversal surface is said to be `global' if it intersects each
$c$-orbit. All these definitions can be extended without change to
the infinite-dimensional case.

	Similarly, the projections of perennials and of symmetries can
be defined as for finite-dimensional systems. Thus, let $i_0 : \Gamma_0
\rightarrow \Gamma$ be the submanifold injection, and let
$\pi_0:\tilde{\Gamma}\rightarrow\Gamma_0$ be the projection that is
defined by $\pi_0 (p) := \gamma_p\cap\Gamma_0$, where $\gamma_p$ is
the $c$-orbit through the point $p\in\tilde{\Gamma}$. If $o$ is a
perennial, then its projection $o_0 : \Gamma_0\rightarrow R$ is
defined by
\begin{equation}
	o_0 := i^*_0 o = o\circ i_0 = o|_{\Gamma_0}.
\end{equation}
If $\psi : \Gamma\rightarrow\Gamma$ is a symmetry, then $a_0 (\psi )
: \Gamma_0\rightarrow\Gamma_0$ is defined by
\begin{equation}
	a_0 (\psi) := \pi_0\circ\psi|_{\Gamma_0}.
\end{equation}
One can easily show that $i^*_0$ is a Poisson algebra isomorphism, and that
$a_0(\psi)$ is a symmetry of $\Gamma_0$.

	We shall use a special type of transversal surface that is
associated with embeddings as follows. For any spacelike embedding
$X:\Sigma\rightarrow{\cal M}$, define the subset
$\Gamma_X\subset\tilde{\Gamma}$ by
\begin{equation}
     \Gamma_X := \{(\varphi,\pi,X)\in \tilde{\Gamma}\mid(\varphi,\pi) \in
     \Gamma_\phi\}.
\end{equation}
In the following steps we shall show that $\Gamma_X$ is a
transversal surface.
\begin{enumerate}
\item The subset $\Gamma_X$ can be considered as the image of the map
     $i_X :\Gamma_\phi\rightarrow \Gamma$ defined by
      $i_X(\varphi,\pi):=(\varphi,\pi,X,P)$ where $P$ is given by
	Eq.\ (\ref{cst1}). Note that ${\text{D}}i_X|_{(\varphi,\pi)}$ is
     given by
     \begin{equation}
          {\text{D}}i_X|_{(\varphi,\pi)} (\Phi,\Pi,V,W)) =
          (\Phi,\Pi,0,W) \in T_{\tilde{C}(\varphi
          ,\pi ,X)}\tilde{\Gamma},
     \end{equation}
     and hence the linear map ${\text{D}}i_X|_{(\varphi,\pi)}$ is
	injective and splits as
     \begin{eqnarray}
          T_{\tilde{C}(\varphi,\pi,X)}\tilde{\Gamma} & = &
          \{(\Phi,\Pi,0,W)\mid(\Phi,\Pi)
          \in T_{(\varphi,\pi)}\Gamma_\phi,\
			W = -{\text{D}}H^\phi(\Phi,\Pi,0)\} \nonumber \\
          &\qquad \times\ &
          \{(0,0,V,W)\mid V\in H^s_0,\ W=-{\text{D}}H^\phi(0,0,V)\}.
							\label{split}
     \end{eqnarray}
     Hence, $\Gamma_X$ is a smooth submanifold of $\Gamma$.

\item  Any tangent vector to $\Gamma_X$ at $(\varphi,\pi ,X)\in
     \Gamma_\phi\times \cal E$ has the form $(\Phi,\Pi,0)$, where
	$(\Phi,\Pi)\in T_{(\varphi,\pi)}\Gamma_\phi \simeq\Gamma_\phi$. The
	tangent space to $\gamma_{(\varphi,\pi,X)}$ at $(\varphi,\pi,X)$ is
	the space $\Xi_{(\varphi,\pi,X)}$ given by Eq.\ (\ref{Xi}), and the
	only vector $(\Phi,\Pi,V)$ in $\Xi_{(\varphi,\pi,X)}$ with $V = 0$
	is the zero vector.  Thus the condition for transversality is
	satisfied.

\item Any $c$-orbit $\gamma_\phi$ intersects $\Gamma_X$, and the point of
     intersection is $(\varphi,\pi,X)$ where $(\varphi,\pi)$ is the
     (unique) Cauchy datum of $\phi$ at $X$. Thus $\Gamma_X$ is a global
     transversal surface.
\end{enumerate}
Note that the injection $i_X$ gives $\Gamma_X$ the structure of a linear
(Fr\'{e}chet) space. Hence we can identify $T_{(\varphi,\pi)}\Gamma_X$ with
$\Gamma_X$ itself.

	The pull-back $\Omega_X$ of $\tilde\Omega$ by $i_X|_{\tilde\Gamma}$
can easily be calculated from Eq.\ (\ref{pulback}) as
\begin{equation}
     \Omega_X((\Phi_1,\Pi_1),(\Phi_2,\Pi_2)) =
		\int_{\Sigma}d^3 x\,(\Phi_2\Pi_1 - \Phi_1\Pi_2).
\end{equation}
This is a constant, weakly non-degenerate form on
$T_{(\varphi ,\pi )}\Gamma_\phi\times T_{(\varphi ,\pi )}\Gamma_\phi$
that can be identified with the following one on $\Gamma_\phi\times
\Gamma_\phi$:
\begin{equation}
     \Omega_X((\varphi_1,\pi_1),(\varphi_2,\pi_2 )) = \int_\Sigma
     d^3x\,(\pi_1\varphi_2 - \varphi_1\pi_2).
\end{equation}
This form can be used to equip $\Gamma_\phi$ with the structure of a
linear, weak-symplectic space.

	Note that the perennial formalism allows for more
general transversal surfaces that are not necessarily associated with
surfaces in space-time. An intriguing---and open---question is if an
explicit example of such a surface can be found and, if so, if it
can be used to construct a quantum field theory on a generic
space-time with no timelike Killing vectors (see later).

	Let us suppose next that there is a one-dimensional group of
isometries $\vartheta (t)$ in the space-time $\cal M$ such that, for
all $t$, $\vartheta(t) (X(\Sigma))\neq X(\Sigma)$ and $\vartheta
(t)$ is generated by an everywhere timelike Killing vector in $\cal
M$. The corresponding one-dimensional group $\{\theta (t)\}$ of
symmetries of $\Gamma$, together with the transversal surface
$\Gamma_X$, form a basis for the construction of an `auxiliary rest
frame' with `time levels' given by $\Gamma_t := \theta (t)\Gamma_X$
and `rest trajectories' given by $\theta$-orbits $\{\theta (t)p\}$,
$p\in \Gamma_0$. Any $c$-orbit $\gamma$ defines a curve,
$t\mapsto\eta_{\gamma}(t) := \Gamma_t\cap \gamma$, and the motion
with respect to the auxiliary rest frame can be defined in a
complete analogy to the finite-dimensional case by comparing the
curve $\eta_{\gamma}(t)$ with the rest trajectories $\theta (t)p$.
`The same measurement at different times' can again be defined as
the set of time shifted perennials $o\rightarrow o_t := s_{\vartheta
(t)}o$, and the construction of the classical Schroedinger or
Heisenberg pictures by means of the projection to $\Gamma_0$ is
straightforward (for details see \cite{timelevels}).

	However, note that the procedure described here differs in one
respect from that described in \cite{timelevels}. Namely, the symmetry
group $\{\theta (t)\}$ we have chosen to generate the time evolution
is {\em not} a subgroup of the first-class canonical group ${\cal
G}_{\text{can}}$ or ${\cal G}_{\text{loc}}$. Thus, the construction
of the quantum mechanical time evolution as given in
\cite{timelevels} has to be generalized.  This will be done in the
next section.

\section{Quantum theory}
\label{Sec:quantum_theory}
In this section, the construction of the quantum theory described in
\cite{timelevels} for finite-dimensional systems will be extended to the
scalar field on a fixed background. The construction uses a representation
of the first-class canonical group $\cal G$ by unitary operators
$R(g)$, $g\in{\cal G}$, on a Hilbert space ${\cal K}$. The generators of
${\cal G}$---the elements of the Lie
algebra ${\cal S}$---are represented by self-adjoint operators on ${\cal
K}$. Then the automorphism $s_{\vartheta (t)}$ of the algebra ${\cal S}$
defines an automorphism $\hat{s}_{\vartheta (t)}$ of the corresponding
operator algebra by the commutative diagram:
\begin{equation}
\begin{array}{ccc}
	{\cal S} & \stackrel{s_{\vartheta (t)}}{\longrightarrow} & {\cal S} \\
	\downarrow R &					   & \downarrow R  \\
   R({\cal S}) & \stackrel{\hat{s}_{\vartheta (t)}}{\longrightarrow} & R({\cal
S})
\end{array}
\label{diagram}
\end{equation}
We arrive at a unitary evolution if we can implement the
automorphism $\hat{s}_{\vartheta (t)}$ by a unitary map $U(t) :
{\cal K} \rightarrow {\cal K}$; that is, $\hat{s}_{\vartheta
(t)}(\hat{O}) = U^{-1}(t)\,\hat{O}\,U(t)$.

	The classical constructions in the previous sections---in
particular, the choice of the algebras of elementary
perennials---were performed in such a way that the rules of the
algebraic or group-theoretical approaches to quantization as
described above lead directly to well-known approaches to the
quantization of a scalar field on a fixed space-time background.
In particular, ${\cal S}_{\text{can}}$ leads to the Segal theory
(for example, see \cite{bongaarts} and \cite{kay}).

	Let us concentrate on ${\cal S}_{\text{can}}$. As was explained
in the previous section, ${\cal S}_{\text{can}}$ is an
infinite-dimensional version of the Heisenberg algebra, and it
determines an abstract infinite-dimensional Heisenberg group ${\cal
G}_{\text{can}}$ (in fact, a `nuclear group', see \cite{G+V}) that
is a central extension of the corresponding symmetry group of the
phase space $\Gamma$ and which acts transitively on the $c$-orbits.
The group ${\cal G}_{\text{can}}$, together with a unitary
representation (which must satisfy certain additional conditions in
order to guarantee the existence of `quantum observables', see
\cite{bongaarts}) is called a `Weyl system' in the literature. If we
apply the theory of Weyl systems to the present case, we can draw
the following conclusions:

\begin{enumerate}
\item The group-theoretical approach to the quantization of an
infinite-dimensional system differs significantly from the
finite-dimensional case in the following respects.  As a rule, a
finite-dimensional Lie group has only relatively few
representations---indeed, the Heisenberg group of $n$ dimensions has
just one (up to unitary equivalence) for any $n$. Many
finite-dimensional canonical groups arise naturally as semi-direct
products in which the `non-abelian' factor is sufficiently large
that there are only a few inequivalent orbits in the dual of the
abelian factor (see \cite{isham}). However, an infinite-dimensional
Heisenberg group has a huge number of non-equivalent
representations.  Many of these have no obvious physical
application, while others have a meaning in relation to external
parameters. For example, in quantum field theory at a finite temperature
each value of the temperature is associated with a particular
representation (and non-zero temperature representations are not
even irreducible).

\item {The most difficult part of the construction of a linear
quantum field theory is therefore the choice of a `physical'
representation.  In the Segal theory, the key object on which such a
choice is based is the time evolution automorphism $s_{\vartheta
(t)}$ of the Heisenberg algebra ${\cal S}_{\text{can}}$. For
example, a cyclic state (generating the representation by a
Gel'fand-Neumark-Segal construction) might be selected using the
Kubo-Martin-Schwinger condition with $\hat{s}_{\vartheta (t)}$.
Another approach based on $s_{\vartheta (t)}$ is described in
\cite{kay}. }
\end{enumerate}

     Thus a new problem arises here, analogous perhaps to the
`Hilbert space problem' of Kucha\v{r}'s classification
\cite{kuch-prehled} of different aspects of the problem of time in
canonical quantum gravity. In fact, quantum field theory in a curved
background has a time problem of its own: most interesting
background space-times do not possess a one-dimensional group of
timelike isometries $\vartheta (t)$ ({\rm i.e.}, there is no
time-like Killing vector), so that $s_{\vartheta (t)}$ is not
available.  However, several methods have been developed for (at
least, partly) bypassing this problem and thereby enabling a number
of interesting questions to be addressed. These methods are not as
mathematically rigorous as those based on a timelike Killing vector
but, nevertheless, they may give some hints about the problem of
time in the full theory of canonical quantum gravity. We shall
consider two different strategies that we shall call the
`scattering approach' and the `algebraic approach' to quantization.
Let us describe how they can be applied in the context of the
perennial formalism.

\subsection{Scattering approach}
\label{Sec:scatter}
The scattering approach to quantization is based on an isolated
symmetry whose domain is a small subset of the phase space $\Gamma$.
Let us consider first an (idealized) example in which $({\cal M},g)$
is a space-time that satisfies the conditions of section
\ref{Sec:phase} and which contains open subsets $U'$ and $U''$ with
the following properties:
\begin{enumerate}
\item Both $U'$ and $U''$ are locally stationary: {\em i.e.}, there
are local flows $\vartheta'(t,X)$ and $\vartheta''(t,X)$ generated
by timelike Killing generators that are defined everywhere on $U'$
and $U''$ respectively.

\item Both $U'$ and $U''$ contain Cauchy hypersurfaces: {\em i.e.},
there are spacelike embeddings $X'$ and $X''$ such that
$X'(\Sigma)\subset U'$ and $X''(\Sigma) \subset U''$.

\item There is an isometry $\vartheta : U'\rightarrow U''$ such that
$X'' = \vartheta\circ X'$.
\end{enumerate}
Finally, let ${\cal S}_{\text{can}}$ denote the algebra of elementary
perennials as discussed in section \ref{Sec:perennials}.

	The local flows $\vartheta'$ and $\vartheta''$ may not induce
global symmetries of $\Gamma$, but they will define perennials $h'$
and $h''$ in some neighbourhood of $\Gamma_{X'}$ and $\Gamma_{X''}$
that correspond to the generators of $\vartheta'$ and $\vartheta''$
respectively. These perennials can be used to construct
representations $(R',{\cal K}')$ and $(R'',{\cal K}'')$ of ${\cal
S}_{\text{can}}$ such that $-h'$ and $-h''$ are represented by
positive, self-adjoint operators $\hat{H}'$ and $\hat{H}''$ (see,
e.g. \cite{wald-haw}).  Following the procedure described in
\cite{patchI}, one could now try to implement the map
$\rho_{X'X''}:\Gamma_{X'}\rightarrow\Gamma_{X''}$, which is a
symplectic diffeomorphism (see section \ref{Sec:phase}), by a
unitary map $U(\rho ):{\cal K}'\rightarrow{\cal K}''$. This would
leave us with only one Hilbert space (a `pasting' of ${\cal K}'$ and
${\cal K}''$).

	However, the literature on the quantum theory of a scalar field
has proceeded in a different direction that can be related to the
Heisenberg picture in the perennial formalism, as described in
\cite{timelevels}. The first observation is that the discrete
isometry $\vartheta$ induces a symmetry $\theta$ that is defined in
some neighbourhood of $\Gamma_{X'}$ in $\Gamma$; in turn, $\theta$
determines a well-defined automorphism $s_{\vartheta}: {\cal
S}_{\text{can}}\rightarrow{\cal S}_{\text{can}}$ of the space of
perennials.  Indeed, for this it is sufficient that $\theta$ maps a
globally transversal surface $\Gamma_{X'}$ onto another such
$\Gamma_{X''}$. The $\theta$-shifted perennials are then completely
determined by their values on $\Gamma_{X''}$, and these are given by
the $\theta$-maps of the restrictions of the original perennials to
$\Gamma_{X'}$. Note that that ${\theta}$ is not a symmetry in the
sense of Ref.
\cite{timelevels} (it is not globally defined); we shall refer to
such a map as a `time shift'.

	The next step is to define the map
$\hat{s}'_{\vartheta}:R'({\cal S}_{\text{can}})\rightarrow R'({\cal
S}_{\text{can}})$ by the obvious analogue of the commutative diagram
(\ref{diagram}), and then to see whether or not it can be implemented by a
unitary map $U(\vartheta):{\cal K}'
\rightarrow {\cal K}'$. If the Cauchy hypersurface is compact, this is
always possible \cite{S-matrix}. Thus one can again work with just
a single Hilbert space. The interpretation of the various
mathematical objects is then that $R'({\cal S}_{\text{can}})$
contains the Heisenberg observables at the `time' $\Gamma_{X'}$;
$R'(s_{\vartheta}({\cal S}_{\text{can}}))$ contains those at the
`time' $\Gamma_{X''}$; the elements of ${\cal K}'$ are the
Heisenberg states; and $U(\vartheta)$ is the unitary scattering
matrix.

	If $U(\vartheta)$ does {\em not\/} exist, a Heisenberg-picture
dynamics can still be used to calculate the expectation
values of time-shifted operators that are well-defined in certain
states. For example, in this way one can calculate the number of
particles within a given finite energy range and a finite volume
that are created from the vacuum of ${\cal K}'$ in the region between
$X'(\Sigma)$ and $X''(\Sigma)$, even though the {\em total\/} number
of created particles diverges.

	Note that the scattering approach will work even if there is
only the `rudiments' of a symmetry, but it will give only
information on what comes `out' if we let something go `in'; what
happens `inside' remains quite undertermined.

\subsection{The Hawking effect}
An example of the scattering approach is the calculation of the
Hawking effect \cite{wald-haw}. In this section, we shall
reformulate this calculation in terms of the perennial formalism.
Our motivation is {\em not} to present a new and conceptually better
derivation of the effect but rather to use this model of the scalar
field on a black-hole background to suggest a possible meaning of a time
shift that does not preserve the domains of transversal surfaces.

	In a general system, a transversal surface $\Gamma_X$ will not
be global ({\em i.e.}, it will not cut {\em all\/} the $c$-orbits
transversally), and the time shift that is available will not
preserve the domains of these surfaces (the domain ${\cal
D}(\Gamma_1)$ of a transversal surface $\Gamma_1$ is the subset of
$\tilde{\Gamma}$ such that the $c$-orbit through any point of ${\cal
D}(\Gamma_1 )$ intersects $\Gamma_1$, see
\cite{timelevels}).

	For example, the toy models studied in \cite{patchI} and
\cite{patchII} do not possess global transversal surfaces, but there
are some that are almost global in the sense that the closure of the
domain contains the whole constraint surface $\tilde{\Gamma}$. The
results of \cite{patchI} suggest that there is no difference between
global and almost global surfaces as far as the quantum theory is
considered.  In \cite{patchII}, two almost global surfaces
$\Gamma_1$ and $\Gamma_2$ were chosen, each with two components,
$\Gamma^{\pm}_1$ and $\Gamma^{\pm}_2$ respectively, giving a total
of four, closed transversal surfaces. There is a discrete symmetry
$\theta$ that maps $\Gamma^+_1$ onto $\Gamma^+_2$, but ${\cal
D}(\Gamma^+_1)\neq {\cal D}(\Gamma^+_2)$. The Hilbert spaces ${\cal
K}_1$ and ${\cal K}_2$ corresponding to the transversal surfaces
$\Gamma^+_1$ and $\Gamma^+_2$ were constructed, and a unitary map
$U(\vartheta ) : {\cal K}_1
\rightarrow {\cal K}_2$ of these Hilbert spaces
was found that corresponds to the classical map $\theta$. However,
the pasting map $\rho :\Gamma^+_1\rightarrow\Gamma^+_2$ is defined
(by the $c$-orbits, see \cite{patchI}) only between some proper
subsets of $\Gamma^+_1$ and $\Gamma^+_2$; the corresponding map
$U(\rho )$ is defined only on a proper subspace ${\cal
K}_{12}\subset {\cal K}_1$, and ${\cal K}_{21} = U(\rho ){\cal
K}_{12}$ is a proper subset of ${\cal K}_2$. Although the map
$U(\rho ) : {\cal K}_{12}
\rightarrow {\cal K}_{21}$ itself is unitary, the corresponding time
evolution operator $U(\rho )\circ U^{-1}(\vartheta ):{\cal
K}_1\rightarrow{\cal K}_1$ is defined only on the subspace ${\cal
K}_{12}$ and so it is not a unitary operator on ${\cal K}_1$. This
leads to a time evolution that can change the norms of states, the
only physical interpretation of which is that the system can be
`lost' or `found' during the time evolution---this can in fact
happen already in the classical theory of this (bizarre) system.

	A similar situation can arise in the context of quantum field
theory on a curved background. Suppose first that there is a Cauchy
surface $\Sigma$ that consists of two components $\Sigma_1$ and
$\Sigma_2$, so that $\Sigma = \Sigma_1 \cup \Sigma_2$.  Then both
$\Sigma_1$ and $\Sigma_2$ are closed surfaces in $\cal M$, and the
space $\Gamma_{\phi}$ of Cauchy data on $\Sigma$ splits into the
direct sum of $\Gamma_{\phi 1}$ and $\Gamma_{\phi 2}$, where
\begin{equation}
     \Gamma_{\phi i} = \{(\varphi ,\pi) \in\Gamma_{\phi}\mid
     {\text{supp}}(\varphi ,\pi)\subset\Sigma_i\},
\end{equation}
$i = 1,2$. Both $\Gamma_{\phi 1}$ and $\Gamma_{\phi 2}$ are
Fr\'{e}chet spaces, and $\Gamma_{\phi} = \Gamma_{\phi
1}\otimes\Gamma_{\phi 2}$. For a given spacelike embedding
$X : \Sigma_1\cup\Sigma_2\rightarrow{\cal M}$, a pair of transversal
surfaces $\Gamma_{Xi}$, $i=1,2$, can be defined by
\begin{equation}
     \Gamma_{Xi} := \{(\varphi ,\pi ,X)\in \tilde{\Gamma}\mid
     {\text{supp}}(\varphi ,\pi)\subset\Sigma_i\}.
\end{equation}
The proof that $\Gamma_{Xi}$ is transversal is analogous
to that for $\Gamma_X$ in section \ref{Sec:perennials}, with the
relation Eq.\ (\ref{split}) being replaced by
\begin{eqnarray}
	T_{\tilde{C}(\varphi ,\pi ,X)}\tilde{\Gamma} & = &
     	\{(\Phi,\Pi,0,W)\mid(\Phi,\Pi)\in
  		\Gamma_{\phi 1},W = -{\text{D}}H^{\phi}(\Phi,\Pi,0)\}
	\nonumber \\
     &&\times\ \{(\Phi,\Pi,0,W)\mid(\Phi,\Pi)\in
     	\Gamma_{\phi 2},W= -{\text{D}}H^{\phi}(\Phi,\Pi,0)\}\nonumber \\
     &&\ \times\ \{(0,0,V,W)\mid V\in T_X{\cal E},
	W=-{\text{D}}H^{\phi}(0,0,V)\}.
\end{eqnarray}
Of course, the surface $\Gamma_{X1}$ is not even almost
globally-transversal: we have the relation $\Gamma_X = \Gamma_{X1}
\times \Gamma_{X2}$, and only $\Gamma_X$ is a global transversal surface.

	The Poisson algebra ${\cal P}$ of perennials contains ideals ${\cal
P}_1$ and ${\cal P}_2$ of perennials associated with the transversal
surfaces $\Gamma_{X1}$ and $\Gamma_{X2}$, where
\begin{equation}
	{\cal P}_i := \{o\in {\cal P}|{\text{supp}}\,o\subset {\cal
	D}(\Gamma_{Xi})\}.
\end{equation}
These ideals ${\cal P}_1$ and ${\cal P}_2$ generate ${\cal P}$.
Similarly, the Lie algebra of elementary perennials ${\cal S}$
contains two ideals ${\cal S}_1$ and ${\cal S}_2$ defined by
analogous equations, and ${\cal S}={\cal S}_1\oplus{\cal S}_2$ is a
Lie algebra decomposition. Indeed, each function $o\in{\cal S}$ has
a restriction $o_X$ to $\Gamma_X$, and there are unique
$o_{X1}\in{\cal S}_1$ and $o_{X2}\in{\cal S}_2$ such that $o_X =
o_{X1}+o_{X2}$. Observe that $o_{Xi}$ vanishes at $\Gamma_{Xi}$, so
that we have $\{o_{X1},o_{X2}\} = 0$ as desired.

	Suppose that the physical representations of the algebras ${\cal
S}$, ${\cal S}_1$ and ${\cal S}_2$ on Hilbert spaces ${\cal K}$,
${\cal K}_1$ and ${\cal K}_2$ respectively have the property ${\cal
K} = {\cal K}_1\otimes_s{\cal K}_2$, where `$\otimes_s$' denotes the
symmetrized tensor product. Let $|a\rangle$ be an arbitrary element
of ${\cal K}$. Then it is well-known that there is a density
operator $\hat{a}_1$ in ${\cal K}_1$ such that
\begin{equation}
	\langle a|\hat{o}_1|a\rangle = {\text{tr}}(\hat{a}_1\hat{o}_1 ), \quad
	{\text{ for all }} \hat{o}_1\in L({\cal K}_1 )
\end{equation}
(for example, see \cite{wald-haw}). Suppose finally that there is an isometry
$\vartheta : X'(\Sigma )\rightarrow X(\Sigma_1 )$, where $X'(\Sigma )$ is
a Cauchy hypersurface and $X(\Sigma_1 )$ is defined as above. The
corresponding time shift $\theta : \Gamma_{X'}\rightarrow \Gamma_{X1}$ maps a
global transversal surface onto a non-global one. Let us define
$s_{\vartheta}$ by
\begin{eqnarray*}
	s_{\vartheta}o|_{\Gamma_{X1}} & = & o\circ\theta^{-1}, \\
	s_{\vartheta}o|_{\Gamma_{X2}} & = & 0 \quad
			{\text{ for all }}o\in{\cal S}.
\end{eqnarray*}
We see immediately that $s_{\vartheta}({\cal S}) = {\cal S}_1$.

	Now we can apply the Heisenberg picture method as described in
subsection \ref{Sec:scatter}. The elements of ${\cal K}$ are
considered as Heisenberg states, and the algebra $R({\cal S})$
contains the Heisenberg observables at the `time' $\Gamma_{X'}$ and
$R({\cal S}_1)$ contains those at the `time' $\Gamma_{X1}$. The
result is that the time evolution operator $\hat{s}_{\vartheta}$
maps the algebra $R({\cal S})$ onto its own proper subalgebra
$R({\cal S}_1 )$ so that the representation $R$ of ${\cal S}_1$ is
not irreducible, and a Heisenberg state $|a\rangle$ that is pure
with respect to the algebra $R({\cal S})$ is a mixed state
$\hat{a}_1$ with respect to the time shifted algebra $R({\cal S}_1
)$.

	An example in which $\vartheta$ maps a global transversal
surface onto a non-global one is the Hawking radiation produced by
the spherically-symmetric, asymptotically flat space-time associated
with a collapsing star (see \cite{wald-haw}) (actually, it is a
limiting case of the procedure above; moreover, the assumption must
be made that the theory in \cite{I+H} can be generalized to an
asymptotically-flat spacetime). In this example, the scalar field
is chosen to have a vanishing mass-parameter $m$, and hence the
dynamics is determined completely by the conformal structure of the
space-time $({\cal M},g)$. The past and future null infinities
${\cal I}^+$ and ${\cal I}^-$ are null hypersurfaces in the
conformal completion, $\bar{{\cal M}}$, of $\cal M$. The
hypersurfaces ${\cal I}^-$ and ${\cal I}^+\cup {\cal H}$ are Cauchy
hypersurfaces for a zero rest mass field, where ${\cal H}$ is the
event horizon in $\cal M$ (see \cite{K+W}). They can be considered
as limits of time-like Cauchy hypersurfaces. Both ${\cal H}$ and
${\cal I}^+$ are closed hypersurfaces in $\bar{{\cal M}}$.
Note that $\Sigma$ and $X_{\pm}$ can be chosen such that $X_- (\Sigma) =
{\cal I}^-$ and $X_+ (\Sigma) = {\cal I}^+$. The maps
$\vartheta$, $\vartheta_+$ and $\vartheta_-$ which were (effectively)
used in \cite{wald-haw} can be described by means of the Eddington-Finkelstein
coordinates $(u,r,\alpha ,\beta )$ and $(v,r,\alpha ,\beta )$ in
some neighbourhoods of ${\cal I}^-$ and ${\cal I}^+$ as follows:
$\vartheta_-$ is defined by $(v,\infty ,\alpha ,\beta ) \rightarrow
(v+t, \infty ,\alpha ,\beta )$, $\vartheta_+$ by $(u,\infty
,\alpha ,\beta ) \rightarrow
(u+t, \infty ,\alpha ,\beta )$, and $\vartheta$ by
$u(v,\alpha ,\beta )=v$, $r=\infty$, $dr\rightarrow -dr$, $\alpha
(v,\alpha ,\beta )=\alpha$, and $\beta (v,\alpha ,\beta )=\beta$
(in-coming modes are mapped into out-going ones). This time shift
$\vartheta$ is not uniquely determined because $u$ and $v$
are defined up to an additive constant, but most of the physically
interesting results do not depend on the choice made. In this
situation, the considerations above are applicable, and the result
is again a non-unitary evolution that sends pure states into mixed
states. This time, the normalization of states is preserved: some
information is lost, but the system itself is not.

\subsection{Algebraic approach}
Hilbert spaces play a less direct role in this approach in which the
basic objects are elements of some algebra of local observables on
which states are defined as linear functionals. One can reformulate
the algebraic approach in terms of the perennial formalism using the
algebra of the smeared fields, ${\cal S}_{\text{loc}}$. We shall not
go into detail here, but just sketch the main ideas.

The local observables are polynomials in the smeared field operators
\begin{equation}
     \hat{\kappa}_f,\quad \hat{\kappa}_f\hat{\kappa}_h,\ldots,
\end{equation}
as well as the regularized stress-energy tensor components
$\hat{T}^{\mu\nu}(p)$ at arbitrary points $p$ of the space-time
$\cal M$.  The stress-energy tensor has an immediate physical
interpretation whereas the smeared field operators play only an
auxiliary role.

The states are defined as certain linear functionals on the above
algebra, with the value of such a state $\sigma$ on an operator
$\hat{o}$ having the physical meaning of the expected value.
Attention is restricted to so-called `quasifree Hadamard states',
whose value on any polynomial of the smeared fields is determined by
its value on the following second-order polynomial
\begin{equation}
     G_\sigma(f,h) = \sigma(\hat{\kappa}_f\hat{\kappa}_h +
     \hat{\kappa}_h\hat{\kappa}_f).
\end{equation}
The bilinear form $G_\sigma(f,h)$ has a kernel $G_\sigma(y_1,y_2)$,
so that
\begin{equation}
     G_\sigma (f,h) = \int_{\cal M} \mbox{d}^4 y_1\,\mbox{d}^4 y_2\,
		G_\sigma(y_1,y_2)f(y_1)h(y_2),
\end{equation}
which satisfies the field equation in each argument $y_1$ and $y_2$.
This leads to the crucial observation that the state can be
`calculated' by solving the wave equation.  For a Hadamard state,
the short-distance behaviour of $G_\sigma(y_1,y_2)$ as
$y_1\rightarrow y_2$ is such that the expected value in the state
$\sigma$ of the stress-energy tensor is well-defined and can be
calculated from $G_\sigma(y_1,y_2)$.

	The quasifree Hadamard states do not form a Hilbert space:
neither a scalar product---nor a linear combination---of a pair of them
is well-defined. However, to obtain a physical interpretation of
such a state it is only necessary to calculate the expected value of
the stress-energy tensor, and this is feasible. For more details see
\cite{F+S+W}.

\section{Conclusions}
We have found an intriguing result: canonical quantization of a
system can lead to a non-unitary time evolution. The result has been
derived by a careful analysis of global properties of the physical
phase space: an aspect that has been rather neglected heretofore.
However, more work is necessary to understand the relations between
the global properties and the time evolution in some generality.
Also, the physical interpretation of the time evolution in
parametrized systems needs to be developed further.

	In the course of our calculations we have seen that the central
commandment of the perennial formalism---to work only with
perennials---is in reality not too restrictive since almost
everything can be viewed as a perennial. In particular, the usual
particle variables of the quantum field theory---for example, creation and
annihilation operators---can be considered as perennials (${\cal
S}_{\text{can}}$), as can the more local, smeared fields in ${\cal
S}_{\text{loc}}$.

	However, these and other insights gained in our paper have only
a relative value in so far as their derivation exploited two special
features of our model---the linearity of the field equations, and
the existence of a background space-time. A question left for future
research is if these structure can be replaced with something that
will work in more complicated cases.

\subsection*{Acknowledgements}
P.H. thanks the Theoretical Physics Group at Imperial College for
their hospitality, and acknowledges useful discussions with
A.~Higuchi and C.~Fewster.  Both authors gratefully acknowledge
financial support from the European Network {\em Physical and
Mathematical Aspects of Fundamental Interactions}.



\begin{thebibliography}{99}

\bibitem{isham-rev1} C.~J.~Isham, ``Conceptual and geometrical problems in
     quantum gravity'', in {\em Recent aspects of Quantum Fields},
	eds. H.~Mitter and H.~Gausterer, Springer-Verlag, Berlin,
	pp 123--230 (1992).

\bibitem{isham-rev2} C.~J.~Isham, ``Canonical quantum gravity and the problem
     of time'', in {\em Integrable Systems, Quantum Groups, and Quantum Field
    Theories}, eds. L.~A.~Ibort and M.~A.~Rodriguez, Kluwer Academic
	Publishers, London, pp 157--288 (1993).

\bibitem{kuch-prehled} K.~V.~Kucha\v{r}, ``Time and interpretations of
    quantum gravity'', in {\em Proceedings of the 4th Canadian
	Conference on General Relativity and Relativistic Astrophysics},
	World Scientific, Singapore, pp 211--314 (1992).

\bibitem{dirac} P.~A.~M.~Dirac, {\em Rev.\ Mod.\ Phys.} {\bf 21}
	392--399 (1949).

\bibitem{timelevels} P.~H\'{a}j\'{\i}\v{c}ek,
	{\em J.\ Math.\ Phys.} {\bf 36} 4612--4638 (1995).

\bibitem{isham}
     C.~J.~Isham, ``Topological and global aspects of quantum
	theory'' in {\em Relativity, Groups and Topology II},
    eds. R.~Stora and B.~S.~DeWitt, North-Holland, Amsterdam,
	pp 1061--1290 (1984).

\bibitem{blue-book} A.~Ashtekar, {\em Lectures on Non-Perturbative Quantum
     Gravity}, World Scientific, Singapore (1991).

\bibitem{K45} P.~A.~M.~Dirac, {\em Can.\ J.\ Math.} {\bf 3} 1--14
	(1951).

\bibitem{hyperspace} K.~V.~Kucha\v{r}, {\em J.\ Math.\ Phys.}
	{\bf 17} 777--791 (1976).

\bibitem{K47} K.~V.~Kucha\v{r}, {\em J.\ Math.\ Phys.} {\bf 17}
		801--820 (1976).

\bibitem{I+K} C.~J.~Isham and K.~V.~Kucha\v{r}, {\em Ann.\ Phys.} {\bf 164}
      288--315 (1985).

\bibitem{K33} K.~V.~Kucha\v{r}, ``Canonical methods of
	quantization'', in {\em Quantum Gravity 2: A Second Oxford
     Symposium}, eds. C.~J.~Isham, R.~Penrose and D.~W.~Sciama,
	Clarendon University Press, Oxford, pp 329--374 (1981).

\bibitem{patchI} P.~H\'{a}j\'{\i}\v{c}ek, A.~Higuchi and J.~Tolar,
	{\em Jour.\ Math.\ Phys.}, {\bf 36} 4639--4666 (1995).

\bibitem{patchII} P.~H\'{a}j\'{\i}\v{c}ek, in preparation (1995).

\bibitem{I+H} P.~H\'{a}j\'{\i}\v{c}ek and C.~J.~Isham, ``The
	symplectic geometry of a parametrized scalar field on a curved
	background''. Imperial College preprint IMPERIAL/TP/95-96/1,
	gr-qc/9510028 (1995).

\bibitem{C+M} P.~R.~Chernov and J.~E.~Marsden, {\em Properties of
     Infinite Dimensional Hamiltonian Systems}, Lecture Notes in
     Mathematics, eds. A.~Dold and B.~Eckmann, Springer-Verlag, Berlin
	(1974).

\bibitem{C-B} Y.~Choquet-Bruhat, ``Hyperbolic Partial Differential Equations on
	a Manifold'', in {\em Battelle Rencontres. 1967 Lectures
	in Mathematics and Physics}, eds. Cecile M.~DeWitt
	and John A.~Wheeler, Benjamin, New York, pp 84--106 (1968).

\bibitem{dewitt} B.~S.~DeWitt, {\em Phys.\ Reports} {\bf 19C} 295--357 (1975);
	{\em Dynamical Theory of Groups and Fields}, Gordon and Breach,
	New York (1965).

\bibitem{bongaarts} P.~J.~M.~Bongaarts, ``Linear Fields According to
	I.~E.~Segal'', in {\em Mathematics of
     Contemporary Physics}, ed. R.~F.~Streater, Academic Press, London,
     pp 187--208 (1972).

\bibitem{kay} B.~S.~Kay, {\em Comm.\ Math.\ Phys.} {\bf 62} 55--70 (1978).

\bibitem{G+V} I.~M.~Gel'fand and N.~J.~Vilenkin, {\em Generalized
     Functions IV}, Academic Press, New York (1964).

\bibitem{wald-haw} R.~M.~Wald, {\em Comm.\ Math.\ Phys.}
	{\bf 45} 9--34 (1976).

\bibitem{S-matrix} R.~M.~Wald, {\em Ann.\ Phys.} {\bf 118} 490--510 (1979).

\bibitem{K+W} B.~S.~Kay and R.~M.~Wald, {\em Phys.\ Reports}
	 {\bf 207} 49--136 (1991).

\bibitem{F+S+W} S. A. Fulling, M. Sweeny and R. M. Wald, Commun. Math.
     Phys. {\bf 63} 257--264 (1978).

\end{thebibliography}
\end{document}